\documentclass[a4paper,fleqn]{cas-dc}
\usepackage[]{natbib}
\usepackage[]{graphicx}

\newcommand{ \St}{ {\rm St} }
\newcommand{\red}[1]{\color{red}{#1}}
\newcommand{\blue}[1]{\color{blue}{#1}}
\newcommand{\green}[1]{\color{ForestGreen}{#1}}

\begin{document}
\let\WriteBookmarks\relax
\def\floatpagepagefraction{1}
\def\textpagefraction{.001}

\shorttitle{rim accretion of porous aggregate}
\shortauthors{Matsumoto et al.}

\title[mode = title]{Formation of rims around chondrules via porous aggregate accretion}

\author[1,2]{Yuji Matsumoto}[orcid=0000-0002-2383-1216]
\cormark[1]
\cortext[1]{Corresponding author}
\ead{yuji.matsumoto@nao.ac.jp}

\author[3]{Yasuhiro Hasegawa}[]
\author[4]{Nozomi Matsuda}[]
\author[4]{Ming-Chang Liu}[]
\address[1]{Center for Computational Astrophysics, National Astronomical Observatory of Japan, Osawa, Mitaka, Tokyo, 181-8588, Japan}
\address[2]{Institute of Astronomy and Astrophysics, Academia Sinica, No.1, Sec. 4, Roosevelt Rd, Taipei 10617, Taiwan}
\address[3]{Jet Propulsion Laboratory, California Institute of Technology, Pasadena, CA 91109, USA}
\address[4]{Department of Earth, Planetary, and Space Sciences, University of California, Los Angeles, Los Angeles, CA}

\begin{abstract}
Chondrules are often surrounded by fine-grained rims or igneous rims.
The properties of these rims reflect their formation histories.
While the formation of fine-grained rims is modeled by the accretion of dust grains onto chondrules, the accretion should be followed by the growth of dust grains due to the shorter growth timescale than the accretion.
In this paper, we investigate the formation of rims, taking into account the growth of porous dust aggregates.
We estimate the rim thickness as a function of the chondrule fraction at a time when dust aggregate accretion onto chondrules is switched to collisions between these chondrules.
Our estimations are consistent with the measured thicknesses of fine-grained rims in ordinary chondrites.
However, those of igneous rims are thicker than our estimations.
The thickness of igneous rims would be enlarged in remelting events.
\end{abstract}


\maketitle

\section{Introduction} \label{sec:intro}

Understanding the origin and evolution of the solar system is one long-standing problem.
Meteoritic data provide unique constraints on the physical and chemical environments of the solar nebula and on the formation of terrestrial planets and small bodies, such as asteroids.

Chondrules are one of the main components of chondrites and contain key information about solar system formation \citep[e.g.,][]{Scott2007}.
They are roughly millimeter-sized particles,
and some of them are coated with dust grains, which are called rims.
There are two types of rims: fine-grained rims \citep[FGRs, e.g.,][]{King_TVV&King_EA1981} and coarse-grained igneous rims \citep[e.g.,][]{Rubin1984,Matsuda+2019}.
One plausible explanation for the formation of FGRs is the accretion of tiny dust grains by chondrules in the solar nebula \citep[e.g., ][]{Morfill+1998,Liffman2019}.
After FGRs formed, some of them may be transformed into igneous rims by remelting events \citep[][]{Rubin2010}.
Rims and their structure recorded the history of dust accretion and remelting events that the host chondrules experienced in the first few million years of the solar system's history.

Studies of FGRs have revealed that the thickness of rims ($a_{\rm rim}$) increases linearly as the radius of chondrules ($a_{\rm ch}$) increases:
\begin{eqnarray}
	\label{eq:a_rim_obs}
	a_{\rm rim} \approx K a_{\rm ch} +c,
\end{eqnarray}
where $K$ and $c$ range from 0.02 to 0.42 and from 8~$\mu$m to 29~$\mu$m, respectively \citep[e.g.,][{see also Table \ref{table:data_Kc}}]{Metzler+1992,Sears+1993, Friend+2018, Hanna&Ketcham2018}
\footnote{
The linear relation is confirmed in carbonaceous chondrites, especially CM chondrites. 
\cite{Bigolski2017_PhD} showed that the correlation is weak in the Semarkona chondrules and less evident in those found in the unequilibrated ordinary chondrites.
}.

Theoretical estimations of the FGR thickness reproduce the aforementioned relationship between $a_{\rm rim}$ and $a_{\rm ch}$.
\cite{Morfill+1998} considered that accretion of FGRs in a solar nebula is characterized by the local mass densities of chondrule and dust, which are given by $\rho_{\rm ch}$ and $\rho_{\rm d}$, respectively, and derived
\begin{eqnarray}
	\label{eq:a_rim_Morfill}
	a_{\rm rim} = a_{\rm ch} \left[ \left( \frac{\rho_{\rm ch}+\rho_{\rm d} }{ \rho_{\rm ch} + \rho_{\rm d} \exp{ (-t/\tau_{\rm acc} )}} \right)^{1/3} -1\right].
\end{eqnarray}
In the above equation, $\tau_{\rm acc}$ is the typical accretion timescale and can be written as
\begin{eqnarray}
		\tau_{\rm acc} = \frac{\rho_{\rm d}}{\rho_{\rm ch} + \rho_{\rm d} } \frac{m_{\rm ch} }{ Q \rho_{\rm d} \sigma \Delta v },
\end{eqnarray}
where $m_{\rm ch}$ is a chondrule mass, $Q$ is the sticking efficiency, $\sigma$ is the collisional cross-section, and $\Delta v$ is the relative velocity between dust and chondrules.
Equation (\ref{eq:a_rim_Morfill}) is effectively the reflection of mass conservation of the total dust mass around chondrules in the nebula.
Equation (\ref{eq:a_rim_Morfill}) indicates that when $t=0$, no dust grains are accreted, and hence $a_{\rm rim}=0$.
After $t\gg \tau_{\rm acc}$, almost all dust grains are accreted, and the rim thickness becomes
\begin{equation}
\label{eq:chi_local}
a_{\rm rim}=a_{\rm ch}( \chi_{\rm local}^{-1/3} -1 ),
\end{equation}
where $\chi_{\rm local}=\rho_{\rm ch}/(\rho_{\rm ch}+\rho_{\rm d})$ is the initial value of the local chondrule mass fraction.
The $K$ values of $0.1-0.42$ yield the values of $\chi_{\rm local}$ ranging from 0.35 to 0.75.

The above estimation has been improved by several studies, where the more realistic treatments of turbulence, the sticking efficiency, the internal structure of rims, and/or erosive collisions were adopted \citep[][]{Cuzzi+2004,Carballido2011, Xiang_C+2019,Xiang_C+2019b,Liffman2019}.
However, such improvements still do not allow for accurate estimation of the rim thickness.
This is because, as shown by the previous studies \citep{Ormel+2008, Arakawa2017,Matsumoto+2019}, the growth timescale of dust grains is shorter than the accretion timescale of dust grains onto chondrules.
Moreover, when the internal density evolution of growing dust aggregates is properly taken into account, dust growth is further accelerated; dust aggregates initially grow via hit-and-stick collisions, which decreases their bulk densities \citep[e.g.,][]{Blum2004, Wada+2008} and increases their collisional cross-sections.
In such a picture of dust growth, the rim accretion onto chondrules is expected to be more efficient.
Also, the rim thickness should become a function of the chondrule size, chondrule fraction, and disk parameters (c.f., Equation (\ref{eq:a_rim_Morfill})).

In this paper, we adopt a porous dust growth model and estimate the thickness of rims around chondrules.
We compare our results to the thicknesses of both FGRs and igneous rims to consider whether these rim sizes can be explained by accretion.
Our estimation and results are summarized in Section \ref{sec:rim}.
In Section \ref{sec:dis}, we present the conclusion of this work and discuss the formation processes of FGRs and igneous rims.

\section{Estimation of rim thickness} \label{sec:rim}

\subsection{Overview}\label{sect:model}

We consider a local region in a gas disk where chondrules and dust grains co-exist.
The growth of these chondrules and dust grains may be modeled as follows \citep[][]{Arakawa2017,Matsumoto+2019}:
(i) dust grains grow via their collisions, forming fluffy dust aggregates;
(ii) these dust aggregates then accrete onto chondrules;
and (iii) after chondrules accrete enough amount of these dust aggregates, they begin to collide with each other.
Hereafter, chondrules that accreted dust aggregates are referred to as compound aggregates.
The accreted dust particles are regarded as the source material of rims.
Hence, the rim thickness around chondrules can be estimated by specifying the total mass of accreted dust aggregates at a time when the collision between compound aggregates begins.
We hereafter refer this time as $\tau_{\rm rim}$ in the following discussion\footnote{It should be noted that $\tau_{\rm rim}$ is used only for descriptive purposes;
in our model, dust growth is computed, based on the timescale argument, and hence we do not follow the actual time evolution.
Moreover, our $\tau_{\rm rim}$ and $t$ in Equation (\ref{eq:a_rim_Morfill}) have different meanings since the rim accretion pictures are different.}.

We formulate here the chondrule mass fraction ($\chi$) of a compound aggregate as
\begin{equation}
	\chi = \frac{3 m_{\rm ch}}{
	4\pi \left[ \rho_{\rm rim} \left(a_{\rm rim}+ a_{\rm ch} \right)^3 -\rho_{\rm rim} a_{\rm ch}^3 + \rho_{\rm ch} a_{\rm ch}^3 \right]},
\end{equation}
where $m_{\rm ch}$, $\rho_{\rm ch}$, and $a_{\rm ch}$ are the mass, internal density, and radius of a chondrule, respectively, and $\rho_{\rm rim}$ is the density of the rim.
Under the assumption that dust grains around a chondrule are isotropically distributed \citep[][]{Bland+2011} and the densities of chondrules and rims are $\rho_{\rm ch}/\rho_{\rm rim}\sim1$ in chondrites, the above equation is re-written as
\begin{eqnarray}
	a_{\rm rim}
	&=& a_{\rm ch} \left( \left[ \left(\chi^{-1} -1\right) \frac{\rho_{\rm ch}}{\rho_{\rm rim}}+1 \right]^{1/3} -1 \right)
	\nonumber \\
	&\simeq& a_{\rm ch} (\chi^{-1/3} -1 ).
	\label{eq:estimation_ach}
\end{eqnarray}
One should note that while rims are porous in the accretion stage, 
the assumption that $\rho_{\rm ch}/\rho_{\rm rim}\sim1$ is reasonable in estimating the rim thickness. 
This is because fluffy dust rims are compacted in the subsequent evolution processes: the dynamical compression via high-velocity collisions \citep[e.g.,][]{Wada+2008}; static compression via the ram pressure of the disk gas and via self-gravity \citep[e.g.,][]{Seizinger+2012, Kataoka+2013a}.
The porosities of observed FGRs are up to $\sim$50\% \citep{Daly+2018LPSC, Zanetta+2021}, and the realistic density ratio $\rho_{\rm ch}/\rho_{\rm rim}$ is 1 -- 2.
Although the rim thickness is an increasing function of $\rho_{\rm ch}/\rho_{\rm rim}$, the modest variation of the density ratio does not affect the rim thickness significantly.
When $\chi=$0.1 -- 0.8, the rim thickness for the case that $\rho_{\rm ch}/\rho_{\rm rim}=2$ becomes 1.4 -- 1.9 times thicker than that with $\rho_{\rm ch}/\rho_{\rm rim}=1$. 

One observes that the rim thickness is now expressed by simple functions of both $a_{\rm ch}$ and $\chi$.
As described below, the value of $\chi$ at $\tau_{\rm rim}$ is not necessarily the same as the value of $\chi_{\rm local}$ in Equation (\ref{eq:chi_local}).
Note that in our model, it is idealized that compound aggregates stop the accretion of dust aggregates immediately at $\tau_{\rm rim}$ and start collisions between compound aggregates.
It is more realistic to consider that some of the compound aggregates would continue to accrete dust aggregates, that is, chondrules would acquire thicker rims.
In order to quantify this effect, we also estimate the rim thickness using $\chi_{\rm local}$. 
This estimation assumes that all the chondrules accrete all the surrounding dust aggregates at the comparable rate, and hence the resulting rim thickness takes the maximum value; 
if a small number of chondrules accrete all dust aggregates, then the thicker rim can be formed around these chondrules
\footnote{
	We note that both FGRs and igneous rims are not found around all chondrules \citep[e.g.,][]{Rubin1984, Simon+2018}.
	}.
Rims are, however, stripped in subsequent collisions.
Rim grains are lost in collisions between chondrules with rims \citep[][]{Umstatter+2019}, and high-velocity collisions induce chondrule ejection from aggregates \citep[][]{Arakawa2017}.
We ignore these subsequent events and give an upper limit estimation of rim thicknesses.

\subsection{Model calculations}

We perform simulations of the growth of chondrules and dust in the solar nebula to estimate $a_{\rm rim}$ using our dust growth model \citep{Matsumoto+2019}.
Our model computes $\chi$ as a function of $a_{\rm ch}$.

A complete explanation of our model can be found in \cite{Matsumoto+2019}, and hence only a short description is provided here.
Following the model, we assume that the growth stage is determined by the shortest timescale among the growth timescale of dust aggregates, the accretion timescale of dust aggregates onto compound aggregates (or chondrules), and the growth timescale of compound aggregates by their collisions.
The growth and accretion timescales are computed, using the disk and dust growth models described below.

We adopt a power-law disk model similar to the Minimum Mass Solar Nebula model \citep{Hayashi1981} to describe the solar nebula.
The surface density of the gas is given by
$\Sigma_{\rm g} = 2400\mbox{~g~cm}^{-2} \left( r/1~\mbox{au} \right)^{-3/2}$
\citep[e.g.,][]{Ida&Lin2004}.
The temperature is $T=280~\mbox{K} (r/1~\mbox{au})^{-1/2}$.
We consider a local region at 2~au where the accretion of dust onto chondrules took place.
The size of a chondrule ($a_{\rm ch}$) is a free parameter in this work.
The (initial) radius of a dust grain is assumed to be 0.25~$\mu$m.
The surface densities of chondrules and dust are given by
\begin{eqnarray}
	\Sigma_{\rm ch} &=& 10\mbox{~g~cm}^{-2} \chi_{\rm local} \left( \frac{r}{1~\mbox{au}} \right)^{-3/2},\\
	\Sigma_{\rm d} &=& 10\mbox{~g~cm}^{-2} (1-\chi_{\rm local}) \left( \frac{r}{1~\mbox{au}} \right)^{-3/2},
\end{eqnarray}
respectively.
The motions of chondrules, dust, dust aggregates and compound aggregates are governed by Brownian motion and turbulence \citep{Ormel&Cuzzi2007}.
Their (relative) velocities induced by the turbulence are expressed as a function of the turbulent $\alpha$, which is assumed to be $10^{-4}$, according to the paleomagnetic measurements of chondrules \citep[][]{Fu_R+2014, Hasegawa+2016b}.
Their Stokes numbers (St), which are the dimensionless stopping times, are $\St\lesssim \alpha \ll 1$.
These motions are well coupled with the gas since we consider the early growth phase,
where dust aggregates and compound aggregates are still small.
Their vertical scale heights are therefore equal to that of gas \citep[][]{Youdin&Lithwick2007}, which is given as $h_{\rm g}=c_{\rm s}/\Omega_{\rm K}$, where $c_{\rm s}$ is the sound speed and $\Omega_{\rm K}$ is the Keplerian frequency.
These small $\St$ aggregates undergo low velocity ($\lesssim 10\mbox{~cm~s}^{-1}$) collisions.
Comparing these collision velocities to the fragmentation velocity \citep[][]{Wada+2009,Arakawa2017}, these collisions are classified as merging events.
We, therefore, assume the perfect sticking, while lower sticking efficiencies are assumed in previous studies \citep[][]{Cuzzi+2004, Carballido2011}.
Realistically, the sticking efficiency depends on the internal structure of aggregates and is different in each collision.

Our previous work confirmed that the rim thickness does not strongly depend on disk parameters if all collisions are merging events.
This is because the rim thickness is given by the ratio of the growth timescale between the compound aggregates to the accretion timescale.
These timescales are given by their masses divided by their mass densities, collisional cross-sections, and their collision velocities.
In these quantities, only the collision velocities depend on the disk parameters.
The timescale ratio is not affected by most disk parameters when the collision velocities between any aggregates are given by the turbulence.

\subsection{Cross-section bias in the rim thickness measurements} \label{sect:bias}

\begin{figure}
	\centering
	\includegraphics[scale=.75]{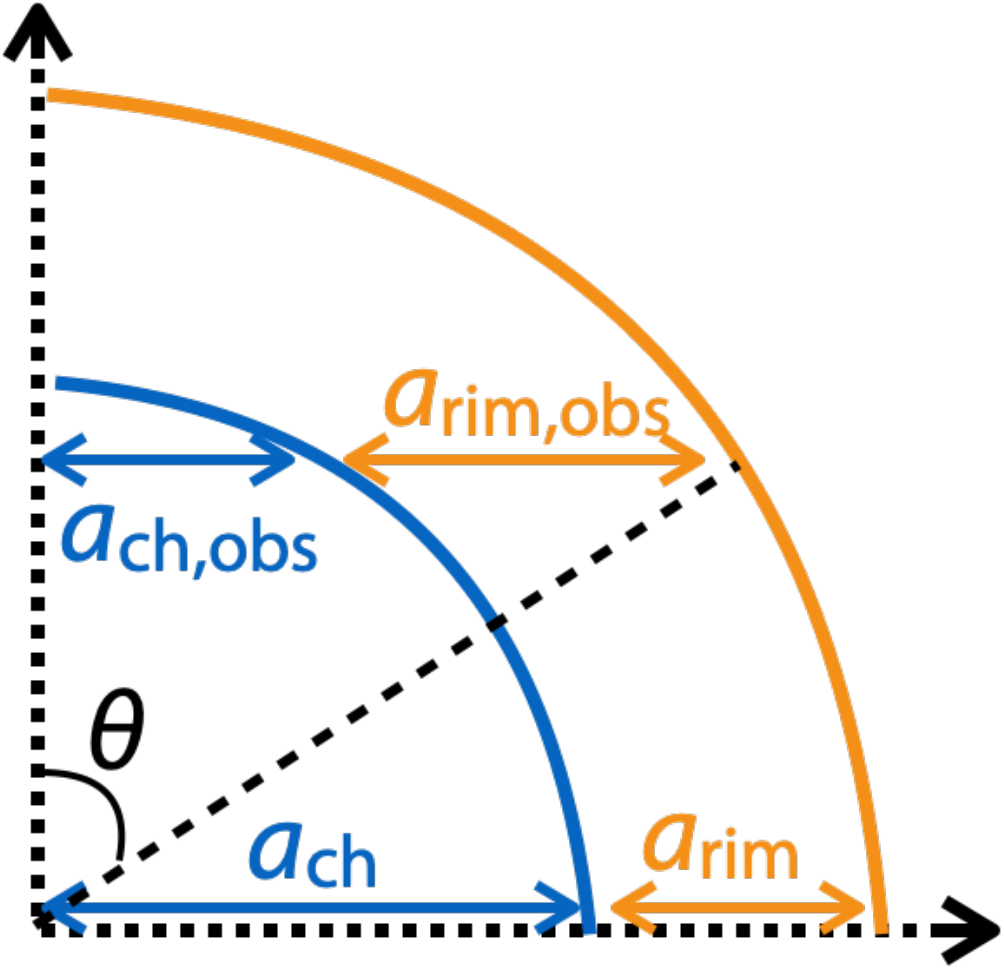}
	\caption{
		Schematic sketch of the cross-section bias.
		It is assumed that a spherical chondrule with the radius of $a_{\rm ch}$ has an isotropic rim whose thickness is $a_{\rm rim}$.
		When this chondrule is cut at a horizontal plane given by an angle $\theta$, the observed chondrule is $a_{\rm ch,obs}$ in radius and has a rim with the thickness of $a_{\rm rim, obs}$.
		}
	\label{Fig:section}
\end{figure}

The size estimates of chondrules and of the surrounding rims are obtained mainly from measurements of two-dimensional cross-sections of meteorites.
It is known that the rim thickness is overestimated due to the cross-section bias.
We here consider this bias in the following simple way.

Figure \ref{Fig:section} shows the schematic picture of our estimation of the cross-section bias.
We assume that chondrules are spherical and homogeneously surrounded by rims.
When we observe a chondrule with a rim at a cutting horizontal plane that forms an angle $\theta$ with the vertical axis, both the observed chondrules radius ($a_{\rm ch,obs}$) and rim thickness ($a_{\rm rim, obs}$) are expressed as a function of $\theta$.
Due to curvatures, the chondrule radius is underestimated and the rim thickness is overestimated as $\theta$ becomes smaller.
A possible range of $a_{\rm ch,obs}$ is between 0 and $a_{\rm ch}$.
We take an average on $a_{\rm ch,obs}$ and $a_{\rm rim, obs}$ in the range of $ \cos^{-1}{(a_{\rm ch}/(a_{\rm ch}+a_{\rm rim}))} \leq\theta\leq 0.5\pi$.
The averaged values of $a_{\rm ch,obs}$ and $a_{\rm rim, obs}$ are not significantly different from the actual values of $a_{\rm ch}$ and $a_{\rm rim}$ since $a_{\rm ch,obs}$ decreases rapidly at $\theta\simeq\cos^{-1}{(a_{\rm ch}/(a_{\rm ch}+a_{\rm rim}))} $.

\subsection{Results}\label{sect:results}

\begin{figure*}
	\centering
	\begin{minipage}{0.45\hsize}
		\includegraphics[width=\textwidth]{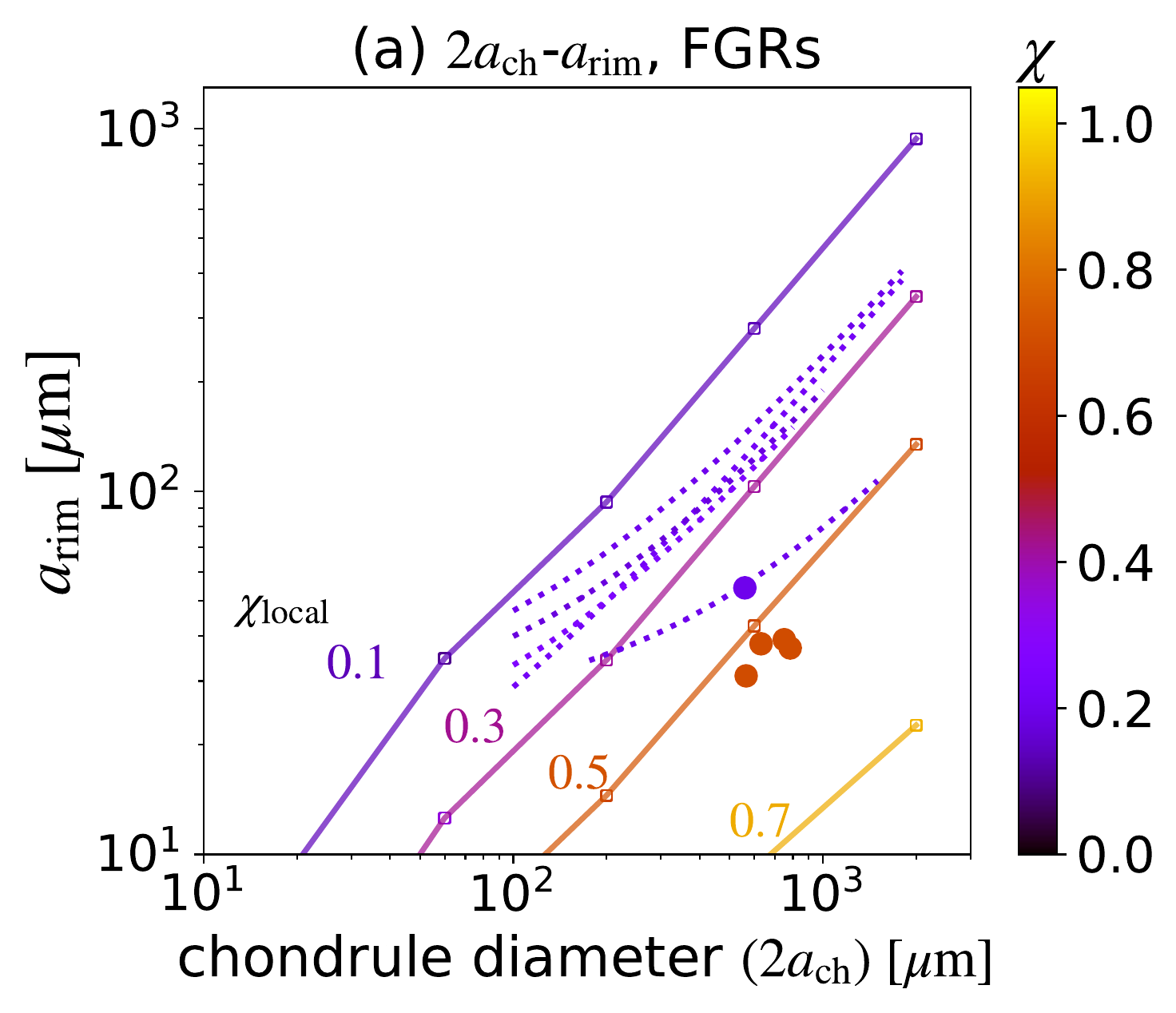}
	\end{minipage}
	\begin{minipage}{0.45\hsize}
		\includegraphics[width=\textwidth]{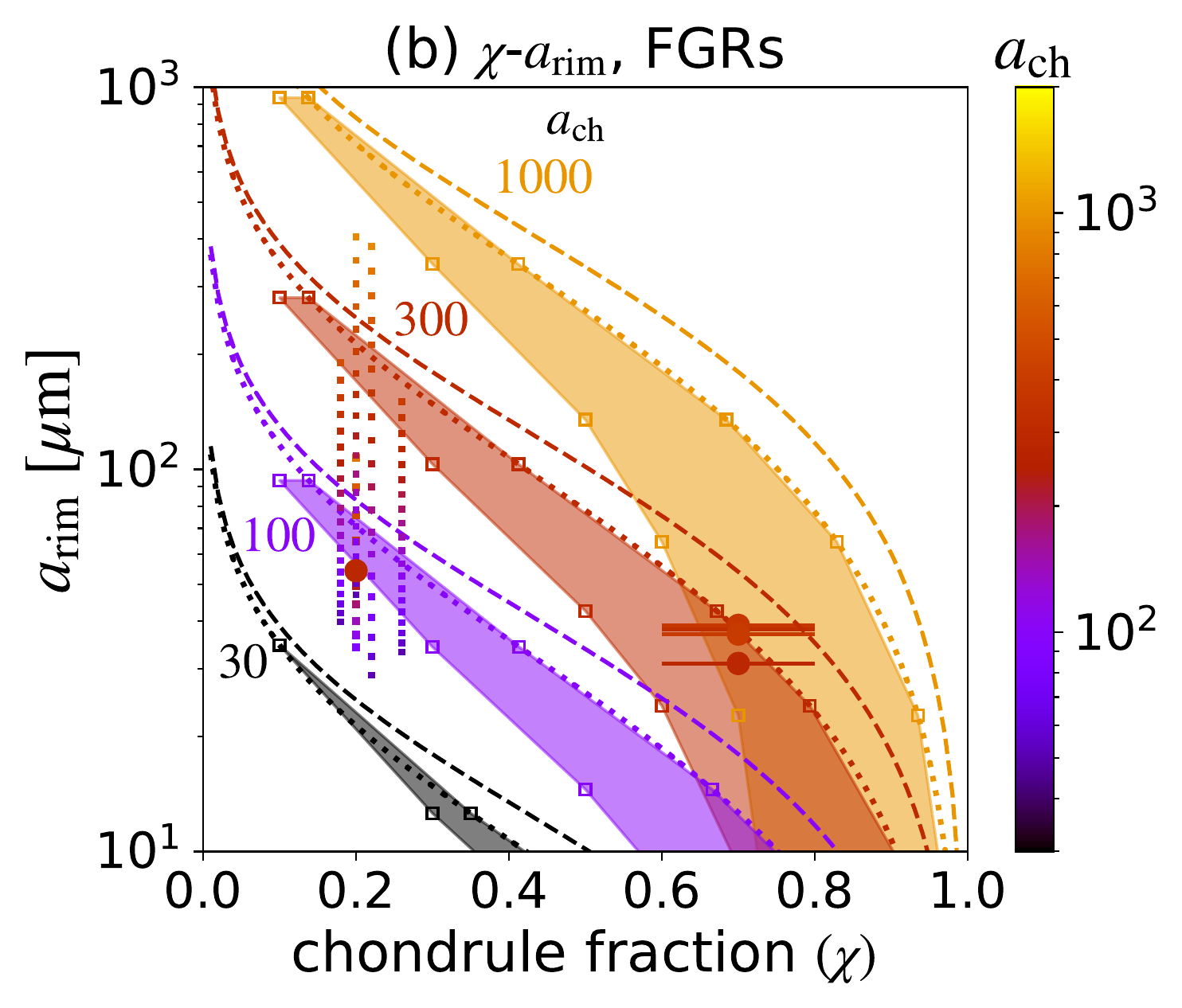}
	\end{minipage}
	\\
	\begin{minipage}{0.45\hsize}
		\includegraphics[width=\textwidth]{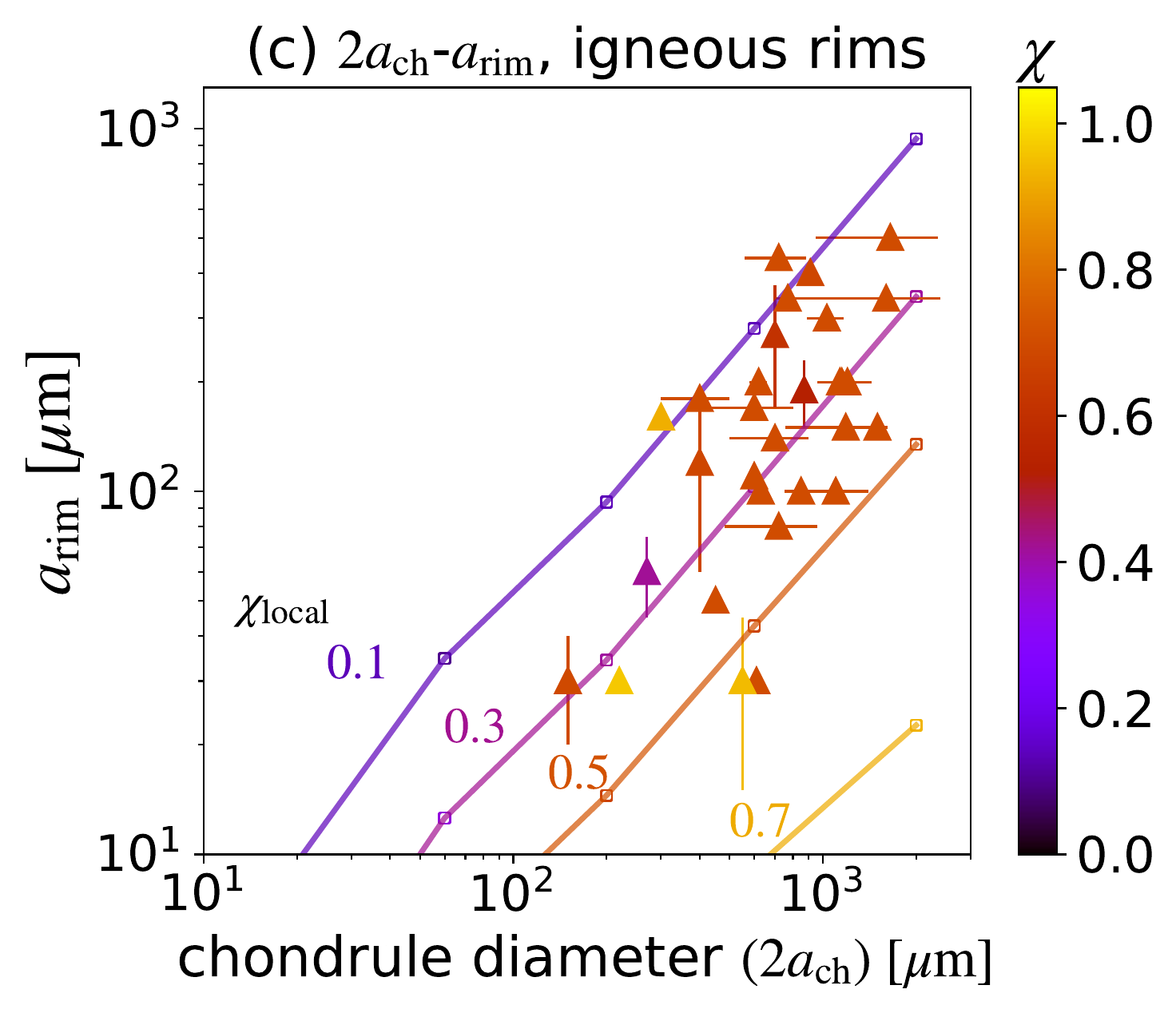}
	\end{minipage}
	\begin{minipage}{0.45\hsize}
		\includegraphics[width=\textwidth]{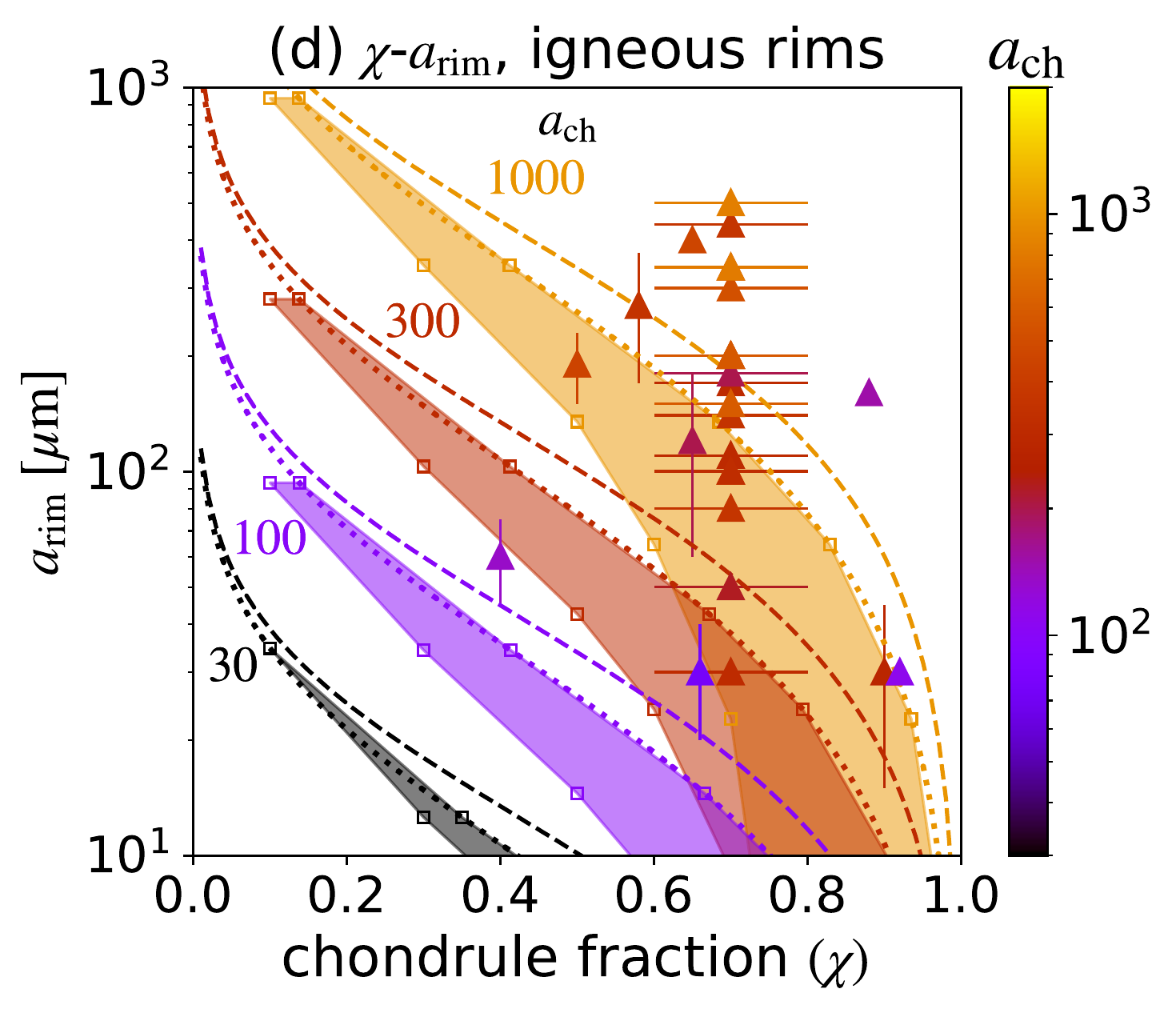}
	\end{minipage}
	\caption{
	Thickness of the rim ($a_{\rm rim}$) is shown as functions of the chondrule diameter ($2a_{\rm ch}$, the left panels) and the chondrule mass fraction ($\chi$, the right panels).
	We show thicknesses of fine-grained dust rims in the upper panels and those of igneous rims in the bottom panels.
	The solid lines with the open squares are our results, which are the same between the upper and bottom panels.
	The filled symbols are measured rims found in the literature (see Table \ref{table:data}, we note that the \cite{Metzler+1992} data in this Table are not plotted since these values are not exact ones).
	When the measured values have ranges, the averaged values are plotted as symbols and the ranges are shown by the solid line segments.
	In the upper panels, we also plot the measured linear relationships between $a_{\rm rim}$ and $2a_{\rm ch}$ (see the dotted lines, Table \ref{table:data_Kc}). 
	The linear relationships by \cite{Metzler+1992} are plotted between $2a_{\rm ch}=100$~$\mu$m and the larger edge of the chondrule diameter in Figure 17 of \cite{Metzler+1992}.
	We note that these measured linear relationships become curved lines on the log-log plot since they have interceptions. 
	\newline
	The details of our results are as follows.
	{\em In the left panels}, the solid lines with the open squares are our simulation results where $\chi_{\rm local}=0.1, 0.3, 0.5$, and 0.7 from the top.
	Different colors represent different chondrule fractions at $\tau_{\rm rim}$ ($\chi$, see the colorbars).
	{\em In the right panels}, the hatched regions bounded by the open squares are given by our numerical results where $a_{\rm ch}=1000, 300, 100$, and 30~$\mu$m from the top.
	These hatched regions are defined by the results with $\chi= \chi_{\rm local}$ (the left edge) and those with $\chi$ at $\tau_{\rm rim}$ (the right edge).
	The dotted lines are obtained from Equation (\ref{eq:estimation_ach}).
	The dashed lines are drawn by applying the cross-section bias to Equation (\ref{eq:estimation_ach}).
	The colors of the symbols represent the values of $a_{\rm ch}$ (see the color bar on the right).
	}
	\label{Fig:a_ch_chi_arim}
\end{figure*}

\begin{figure}
	\centering
	\includegraphics[scale=.75]{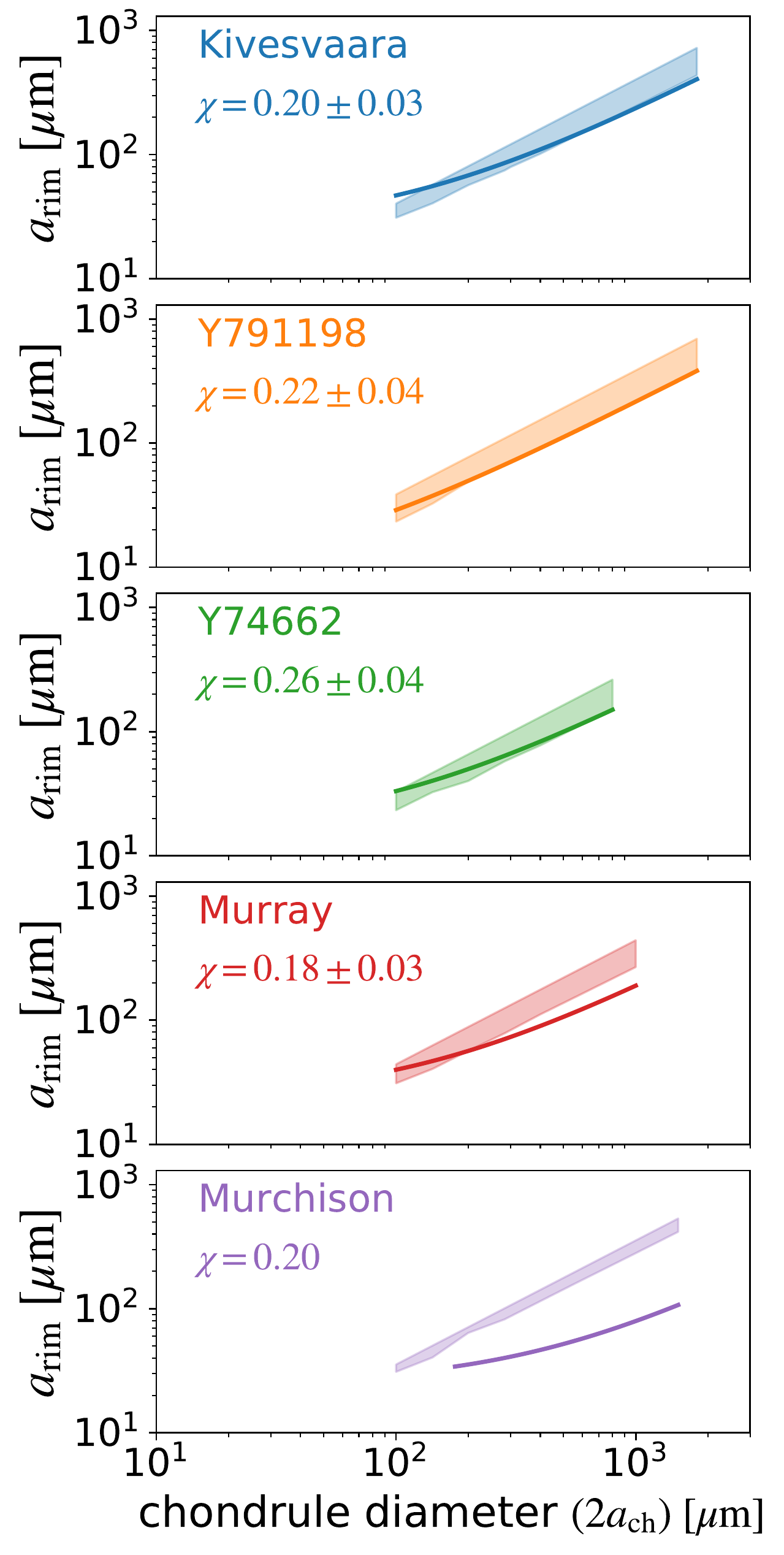}
	\caption{
		Each panel shows the measured linear relationships between the rim thickness ($a_{\rm rim}$) and the chondrule diameter ($2a_{\rm ch}$) of each meteorite (Table \ref{table:data_Kc}) in the solid line.
		The hatched regions are given by the rim thicknesses with $\chi= \chi_{\rm local}$ (the top edge) and those with $\chi$ at $\tau_{\rm rim}$ (the bottom edge).
	}
	\label{Fig:ach_arim_line_mat}
\end{figure}

The dependencies of the thickness of rims ($a_{\rm rim}$) on $a_{\rm ch}$ and $\chi$ are shown in Figure \ref{Fig:a_ch_chi_arim}.
In the left panels, the numerical results show that the thickness of rims is almost proportional to the chondrule diameter ($2a_{\rm ch}$), although there are some tiny fluctuations.
These fluctuations arise because the thickness also depends weakly on $\chi$ ($\chi^{-1/3}$, Equation (\ref{eq:estimation_ach})).

The relations between $a_{\rm rim}$ and $\chi$ are shown in the right panels (see the right edge of the hatched region).
For the comparison purpose, we also plot these thicknesses at $\chi = \chi_{\rm local}$ (see the left edge of the hatched region).
Since chondrules do not accrete all dust particles at the time of $\tau_{\rm rim}$ \citep[][]{Matsumoto+2019}, $\chi$ is larger than $\chi_{\rm local}$.
These two values are approximately the upper and lower limits of the chondrule abundance in finally formed chondrites (Section \ref{sect:model}).
These hatched regions become wider as $\chi_{\rm local}$ increases;
In large $\chi_{\rm local}$ cases, chondrules are the dominant component (over dust particles), and hence compound aggregates quickly undergo collisions between them.
As a result, compound aggregates can keep a high value of $\chi$ (i.e., chondrule-rich).
We also show the measured thicknesses of FGRs in the top panels and igneous rims in the bottom panels.
These observed rim thicknesses of both FGRs and igneous rims tend to increase as the chondrule size increases.
These data are summarized in Appendix (Table \ref{table:data}), where the estimated rim thicknesses are also summarized.

We consider that the measured rim thicknesses agree with our numerical results if the measured values (i.e., $\langle a_{\rm rim}\rangle$, $\langle 2a_{\rm ch}\rangle$, and $\langle \chi \rangle$) are located in the hatched regions (see also Table \ref{table:data} for specific values).
The thicknesses of FGRs in ordinary chondrites \citep{Bigolski2017_PhD} are well reproduced by our results.
The comparisons between our estimation and the observed linear relation are performed in Figure \ref{Fig:ach_arim_line_mat}, where the estimated rim thickness ranges are plotted on the chondrule diameter and rim thickness plane as the hatched regions.
The observed linear relations are consistent with our estimation in CM Y791198 and CM Y74662.
Most of their rim thicknesses are close to the estimated rim thicknesses from $\chi$ at $\tau_{\rm rim}$.
The rim thicknesses become larger than the estimated rim thicknesses from $\chi$ at $\tau_{\rm rim}$ as chondrule diameters decrease due to the interceptions of the observed linear relations.
In CM Kivesvaara, the measured rim thicknesses are consistent with our estimation in most of their chondrule diameter range, although the measured rim thicknesses are slightly larger than the estimated thicknesses around $100~{\rm \mu}$m diameters and slightly smaller than the estimated thicknesses at $1800~{\rm \mu}$m diameter.
However, the rim thicknesses around Murray chondrules are consistent only at $2a_{\rm ch}< 200~{\rm \mu m}$ and those around Murchison chondrules are always thinner than our estimation.

In contrast, the thicknesses of most igneous rims are larger than our results (Figure \ref{Fig:a_ch_chi_arim} and Table \ref{table:data}). 
Only two data points of the rims in ALHA 77034 are consistent with our estimation, and the other rims are thicker than our estimation.
Because our estimation of the maximum rim thickness is based on the $\chi=\chi_{\rm local}$, there are two possible explanations: the cross-section bias; the subsequent acquisition of rim materials by mass transfer.
Large thicknesses of igneous rims can be reproduced by artificially adjusting the cross-section bias;
however, such adjustment makes the apparent chondrule size very small,
that is, 87\% of chondrules should be measured at $<0.4$ times smaller than their actual sizes.
This suggests that igneous rims are not simply formed through dust aggregate accretion since it is unlikely that almost all chondrules are cut at such small $\theta$ values.
Our results thus suggest that igneous rims would be made thicker by the subsequent processes such as remelting events.

\section{Discussion \& Conclusions}\label{sec:dis}

Rims around chondrules are often found in chondrites.
The thicknesses of fine-grained rims in CM chondrites is known to be proportional to the size of its host chondrule.
This linear relation was explained by theoretical studies.
However, the estimation of the thickness of rims in previous studies did not include the growth of dust, although the growth timescale of dust is shorter than the accretion timescale of dust onto chondrules.
We investigate the growth of chondrules and dust and estimate the thickness of the rims.
Our estimation is an upper limit estimation of rim thicknesses since we do not consider the subsequent rim erosion events in aggregate growth processes.
Our estimation agrees with the thicknesses of observed fine-grained rims in ordinary chondrites.
Some of the linear relationships observed in CM chondrites are consistent with our estimation, but the rim thicknesses around CM Murray and CM Murchison are smaller than our estimation.
The chondrules in these CM chondrites would lose their surrounding dust rims in the latter accretion processes such as collisions between chondrules with rims \citep[][]{Umstatter+2019} and collision-induced chondrule ejection \citep[][]{Arakawa2017}.
The thickness of the rims would be affected by the high-velocity collisions and subsequent fragments accretion \citep{Liffman2019}.
In addition, our model considers identical-sized chondrules.
The size distribution of the initial chondrules would affect their rim thicknesses.
These processes should be taken into account to consider the rim thicknesses around CM Murray and CM Murchison.

Besides, the rim thickness of CM chondrite is affected by parent body processes \citep[][]{Sears+1993, Trigo-Rodriguez+2006, Takayama&Tomeoka2012}.
Our results may imply that both the accretion of dust aggregates and the subsequent parent body processes are important to fully understand the rim thickness around CM chondrites.
More rim measurements and future simulations, where the above processes are included, provide more detailed views of the rim formation.

In contrast, the thicknesses of igneous rims are larger than our estimation.
This suggests that igneous rims would not be simply converted from fine-grained rims but get larger in their formation or subsequent processes.
Igneous rims are formed via remelting events, and they may be formed from the materials in chondrule surfaces and their surrounding rims \citep[e.g.,][]{Krot&Wasson1995}.
The accretion or collisions in these remelting events is one possible explanation of why igneous rims are thick.
The accretion of molten aggregates onto surficial melting chondrules would make thicker igneous rims.
These accretion events would be similar to the formation events of compound chondrules \citep[e.g.,][]{Wasson+1995}.
The size evolution of rims in the remelting events would be the key to understanding the thickness of igneous rims.
A more comprehensive study of the thickness of chondrule rims (both fine-grained and igneous) will help improve our modeling and understanding of the formation mechanism of the rims, which shed light on the origin of the solar system.

\section*{Acknowledgments}

We thank John N. Bigolski and the anonymous referee for constructive comments.
Y.H. is supported by the Jet Propulsion Laboratory, California Institute of Technology, under a contract with the National Aeronautics and Space Administration.

\appendix

\section{Observation data}
The measurements available in the literature are summarized in Tables \ref{table:data_Kc} and \ref{table:data}.
In Table \ref{table:data_Kc}, the observed linear relations between chondrule radii and rims are summarized.
In Table \ref{table:data}, we show the averaged values of chondrule diameters ($\langle 2a_{\rm ch}\rangle$), the chondrule fraction ($\langle\chi\rangle$), and the rim thickness ($\langle a_{\rm rim}\rangle$). 
We note that these chondrule fractions are not mass fractions but volume fractions.
Also, in some papers, we calculate chondrule fractions by subtracting matrix fractions from unity.
If the chondrule or matrix fractions are not given, we use chondrule fractions from \cite{Scott&Krot2014}.
The estimated rim thickness is also shown and compared with $\langle a_{\rm rim}\rangle$ in this table.

\begin{table*}[width=.9\textwidth,cols=6,pos=h]
	\caption{
		The values of $K$ and $c$ in Equation (\ref{eq:a_rim_obs}), taken from (1) \cite{Metzler+1992}; (2) \cite{Hanna&Ketcham2018}; (3) \cite{Friend+2018}.
		When the slope values of the diameters ($K_{\rm d}$) are derived in these studies, $K$ is simply estimated as $K=2K_{\rm d}$.
		If the chondrule fractions $\chi$ of the chondrites are not given,
		$\chi$ values are taken from (4) \cite{Scott&Krot2014}.
	}
	\label{table:data_Kc}
	\begin{tabular*}{\tblwidth}{@{} LLLLL@{} }
	\toprule
	$K$	&	$c$ [$\mu$m]	&	$\langle\chi\rangle$~\%	&	Chondrite	&	Reference\\
	\midrule
	0.4214	&	25.8	&	$20\pm3$	&	CM (Kivesvaara)	&	1 \\
	0.4156	&	8.1		&	$22\pm4$	&	CM (Y791198)	&	1 \\
	0.3342	&	16.47	&	$26\pm4$	&	CM (Y74662)		&	1 \\
	0.333	&	23.2	&	$18\pm3$	&	CM (Murray)		&	1 \\
	0.11	&	24.5	&	20			&	CM (Murchison)	&	2,4\\
	0.36	&	--	&	20	&	CM (Jbilet Winselwan)	&	3,4\\
	0.24	&	--	&	20	&	CM (Jbilet Winselwan)	&	3,4\\
	\bottomrule
	\end{tabular*}
\end{table*}
\footnotetext{ungrouped ordinary chondrite}

\begin{table*}[width=.9\textwidth,cols=6,pos=h]
	\caption{
		The data is taken from the following references:
		(1) \cite{Bigolski2017_PhD}; (2) \cite{Scott&Krot2014} (for chondrule fraction $\chi$); (3) \cite{Hanna&Ketcham2018};
		(4) \cite{Krot&Wasson1995}; (5) \cite{Rubin2010}.
		The data taken from (6) \cite{Metzler+1992} is also added for references, but these are not the averaged values. 
		These chondrules diameters are around the maximum values from their Figure 17, and the rim thicknesses are calculated by their linear relationships.
		The estimated rim thicknesses are also compared with the measured thicknesses: 
		The comparison equation is in green when they are consistent; in red when the estimated rim is thinner; in blue when the estimated rim is thicker.
	}
	\label{table:data}
	\begin{tabular*}{\tblwidth}{@{} LLLLLL|L@{} }
	\toprule
	&	$\langle 2a_{\rm ch}\rangle$	[$\mu$m]	&	$\langle\chi\rangle$~\%	&	$\langle a_{\rm rim}\rangle$~[$\mu$m]	&	Chondrite	&	Reference	&	Estimated rim thickness~[$\mu$m]\\
	 \midrule
	 FGR	&	631	&	60 -- 80	&	38	&	LL (Semarkona)	&	1,2 	&	9.8 -- 58.6 \green{($\in \langle a_{\rm rim}\rangle$)} \\
	 FGR	&	750	&	60 -- 80	&	39	&	LL (Watonga)	&	1,2 	&	10.3 -- 69.6 \green{($\in \langle a_{\rm rim}\rangle$)} \\
	 FGR	&	565	&	60 -- 80	&	31	&	LL (Bishunpur)	&	1,2	&	9.3 -- 52.4 \green{($\in \langle a_{\rm rim}\rangle$)} \\
	 FGR	&	784	&	60 -- 80	&	37	&	OC (NWA 5717)	&	1,2	&	10.6 -- 72.8 \green{($\in \langle a_{\rm rim}\rangle$)} \\
	 FGR	&	559.6	&	20	&	54.2	&	CM (Murchison)	&	2,3	&	159.3 -- 198.7 \blue{($>\langle a_{\rm rim}\rangle$)} \\
	 igneous	&	950 -- 1250	&	60 -- 80	&	100	&	L (ALHA 77034)	&	2,4	&	11.0 -- 116.0 \green{($\in \langle a_{\rm rim}\rangle$)} \\
	 igneous	&	600 -- 620	&	60 -- 80	&	30	&	L (ALHA 77034)	&	2,4	&	10.2 -- 57.5 \green{($\in \langle a_{\rm rim}\rangle$)} \\
	 igneous	&	450	&	60 -- 80	&	30 -- 50	&	L (ALHA 77034)	&	2,4	&	8.1 -- 41.8 \red{($\lesssim \langle a_{\rm rim}\rangle$)} \\
	 igneous	&	800 -- 900	&	60 -- 80	&	100	&	L (ALHA 77176)	&	2,4	&	10.7 -- 83.5 \red{($<\langle a_{\rm rim}\rangle$)} \\
	 igneous	&	1100 -- 1200	&	60 -- 80	&	10 -- 200	&	L (ALHA 77176)	&	2,4	&	11.3 -- 111.4 \red{($\lesssim \langle a_{\rm rim}\rangle$)} \\
	 igneous	&	640 -- 800	&	60 -- 80	&	140 -- 440	&	L (ALHA 77260)	&	2,4	&	10.3 -- 74.3 \red{($<\langle a_{\rm rim}\rangle$)} \\
	 igneous	&	600 -- 640	&	60 -- 80	&	200	&	L (ALHA 77260)	&	2,4	&	10.2 -- 59.4 \red{($<\langle a_{\rm rim}\rangle$)} \\
	 igneous	&	600	&	60 -- 80	&	110	&	L (ALHA 77260)	&	2,4	&	10.2 -- 55.7 \red{($<\langle a_{\rm rim}\rangle$)} \\
	 igneous	&	1500	&	60 -- 80	&	75 -- 150	&	L (LBW 85339)	&	2,4	&	11.8 -- 139.2 \red{($\lesssim \langle a_{\rm rim}\rangle$)} \\
	 igneous	&	740 -- 800	&	60 -- 80	&	260 -- 340	&	L (LBW 86158)	&	2,4	&	10.2 -- 74.3 \red{($<\langle a_{\rm rim}\rangle$)} \\
	 igneous	&	1080 -- 1320	&	60 -- 80	&	80 -- 200	&	LL (Bishunpur)	&	2,4	&	11.4 -- 122.5 \red{($<\langle a_{\rm rim}\rangle$)} \\
	 igneous	&	600 -- 800	&	60 -- 80	&	0 -- 140	&	LL (Bishunpur)	&	2,4	&	10.2 -- 74.3 \red{($<\langle a_{\rm rim}\rangle$)} \\
	 igneous	&	500 -- 700	&	60 -- 80	&	50 -- 170	&	LL (Bishunpur)	&	2,4	&	7.6 -- 65.0 \red{($<\langle a_{\rm rim}\rangle$)} \\
	 igneous	&	960 -- 1100	&	60 -- 80	&	30 -- 300	&	LL (Bishunpur)	&	2,4	&	11.2 -- 102.1 \red{($<\langle a_{\rm rim}\rangle$)} \\
	 igneous	&	350 -- 450	&	60 -- 80	&	160 -- 180	&	LL (Bishunpur)	&	2,4	&	7.5 -- 41.8 \red{($<\langle a_{\rm rim}\rangle$)} \\
	 igneous	&	1100 -- 1170	&	60 -- 80	&	60 -- 200	&	LL (Chainpur)	&	2,4	&	11.3 -- 108.6 \red{($\lesssim \langle a_{\rm rim}\rangle$)} \\
	 igneous	&	630	&	60 -- 80	&	100	&	LL (Chainpur)	&	2,4	&	9.8 -- 58.5 \red{($<\langle a_{\rm rim}\rangle$)} \\
	 igneous	&	970 -- 1400	&	60 -- 80	&	100 -- 150	&	LL (Chainpur)	&	2,4	&	11.3 -- 129.9 \red{($\lesssim \langle a_{\rm rim}\rangle$)} \\
	 igneous	&	1200 -- 2000	&	60 -- 80	&	340	&	LL (Krymka)	&	2,4	&	11.7 -- 185.6 \red{($<\langle a_{\rm rim}\rangle$)} \\
	 igneous	&	600 -- 840	&	60 -- 80	&	80	&	LL (Krymka)	&	2,4	&	10.2 -- 78.0 \red{($<\langle a_{\rm rim}\rangle$)} \\
	 igneous	&	1300 -- 2000	&	60 -- 80	&	340 -- 500	&	LL (Krymka)	&	2,4	&	11.7 -- 185.6 \red{($<\langle a_{\rm rim}\rangle$)} \\
	 igneous	&	910	&	65	&	400	&	CV3	&	5	&	35.7 -- 70.3 \red{($<\langle a_{\rm rim}\rangle$)} \\
	 igneous	&	870	&	50	&	190	&	CK3	&	5	&	74.0 -- 113.1 \red{($<\langle a_{\rm rim}\rangle$)} \\
	 igneous	&	700	&	58	&	270	&	CR2	&	5	&	42.2 -- 69.7 \red{($<\langle a_{\rm rim}\rangle$)} \\
	 igneous	&	400	&	65	&	120	&	R3	&	5	&	20.0 -- 30.9 \red{($<\langle a_{\rm rim}\rangle$)} \\
	 igneous	&	270	&	40	&	60	&	CM2	&	5	&	38.4 -- 48.2 \red{($<\langle a_{\rm rim}\rangle$)} \\
	 igneous	&	150	&	66	&	30	&	CO3	&	5	&	10.4 -- 11.1 \red{($<\langle a_{\rm rim}\rangle$)} \\
	 igneous	&	550	&	90	&	30	&	EL3	&	5	&	2.3 -- 9.8 \red{($<\langle a_{\rm rim}\rangle$)} \\
	 igneous	&	220	&	92	&	30	&	EH3	&	5	&	1.6 -- 3.1 \red{($<\langle a_{\rm rim}\rangle$)} \\
	 igneous	&	300	&	88	&	160	&	H	&	5	&	3.1 -- 6.5 \red{($<\langle a_{\rm rim}\rangle$)} \\
	 FGR	&	$\lesssim1800$	&	$20\pm3$	&	$\lesssim 405$	&	CM (Kivesvaara)	&	6	&	$\lesssim434.5$ -- 724.6 \blue{($>\langle a_{\rm rim}\rangle$)} \\
	 FGR	&	$\lesssim1800$	&	$22\pm4$	&	$\lesssim 382$	&	CM (Y791198)	&	6	&	$\lesssim382.0$ -- 694.0 \green{($\in \langle a_{\rm rim}\rangle$)} \\
	 FGR	&	$\lesssim800$	&	$26\pm4$	&	$\lesssim 150$	&	CM (Y74662)		&	6	&	$\lesssim149.3$ -- 262.6 \green{($\in \langle a_{\rm rim}\rangle$)} \\
	 FGR	&	$\lesssim1000$	&	$18\pm3$	&	$\lesssim 190$	&	CM (Murray)		&	6	&	$\lesssim267.8$ -- 441.0 \blue{($>\langle a_{\rm rim}\rangle$)}\\
	\bottomrule
	\end{tabular*}
\end{table*}

\printcredits

\bibliographystyle{cas-model2-names}
\bibliography{bibtex_ym}

\end{document}